# Thermal-emission measurements near room temperature using Fourier-transform infrared spectroscopy

Yuzhe Xiao[1], Alireza Shahsafi[1], Chenghao Wan[1,2], Patrick J. Roney[1], Graham Joe[1], Zhaoning Yu[1,3], Jad Salman[1], and Mikhail A. Kats[1,2,3*]

[1]Department of Electrical and Computer Engineering, University of Wisconsin-Madison, Madison, Wisconsin 53706, USA

[2]Department of Materials Science and Engineering, University of Wisconsin-Madison, Madison, Wisconsin 53706, USA

[3]Department of Physics, University of Wisconsin - Madison, Madison, Wisconsin 53706, USA

*Email address: mkats@wisc.edu

Accurate characterization of thermal emitters can be challenging due to the presence of background thermal emission from components of the experimental setup and the surrounding environment. This is especially true for an emitter operating close to room temperature. Here, we explore characterization of near-room-temperature thermal emitters using Fourier-transform infrared (FTIR) spectroscopy. We find that the thermal background arising from optical components placed between the beam splitter and the detector in an FTIR spectrometer appears as a "negative" contribution to the Fourier-transformed signal, leading to errors in thermal-emission measurements near room temperature. Awareness of this contribution will help properly calibrate low-temperature thermal-emission measurements.

**Introduction**

Thermal emission is a fundamental physical process by which all objects at temperatures above 0 K emit light [1]. By analyzing thermally emitted light, information related to the emitter's material properties [2] and temperature distribution [3] can be extracted. In the past decade, new developments in photonics and nanoscale fabrication have enabled research into engineered thermal emission [4], [5], for applications that include passive radiative cooling [6], efficient incandescent light sources [7], and thermophotovoltaics [8]. Therefore, accurate experimental characterization of thermal emission has become increasingly important. Many thermal-emission measurements are performed via reflection and transmission measurements and the subsequent application of Kirchhoff's law [9], [10], which relates the emissivity of an object in thermal equilibrium to its absorptivity. However, this approach can be difficult for scattering [11] or highly absorbing samples [12]. Furthermore, in systems that are not in thermal equilibrium [13], [14] or are not reciprocal [15], [16], Kirchhoff's law does not apply, and a direct measurement of the emitted light is required. In the infrared region, where most thermal-emission measurements are performed, FTIR spectrometers are more commonly used compared to dispersive instruments due to the so-called "multiplex advantage", which describes the superior signal-to-noise ratio of FTIRs, especially at high resolution [17].

For a significant portion of thermal-emission research and development, such as for thermophotovoltaics, thermal-emission experiments are performed at high temperatures [10], [18], [19], such that the signal dominates over the thermal background emitted by components of the measurement instrument or the room that houses the experiment. However, recently, there has been growing interest in thermal emitters that operate at moderate temperatures [6], [20], [21], where direct-emission measurements can be challenging due to a low signal-to-noise ratio and a relatively large thermal background. Typically, interpreting FTIR-based thermal-emission measurements involves calibrating the FTIR response function and the background [22]–[24]. However, most studies that we have seen assume a response function that is independent of the sample temperature. This assumption turns out to potentially be problematic for cases where thermal emitters are close to room temperature.

In this work, we performed spectral measurements of near-room-temperature thermal emitters using an FTIR spectrometer, and observed a background signal that sometimes appeared to be negative, and a sign inversion of the instrument response function when the sample temperature dropped below room temperature. We found that, in our instrument, these unexpected features arise from the background emission of the room-temperature optical components positioned between the interferometer and the detector. If such background emission is present in an FTIR-based thermal-emission measurement apparatus, special care must be taken to extract quantitatively correct data.

**Experiments**

We used an FTIR spectrometer (Bruker VERTEX 70) to measure thermal-emission spectra from a laboratory blackbody (~500-$\mu$m-tall vertically aligned carbon nanotube (CNT) forest [25]), as a function of temperature [Fig. 1(a)]. The sample was affixed on a temperature-controlled stage mounted inside the FTIR sample compartment, and was rotated by 10° with respect to the beam path to avoid multiple reflections between the sample and the interferometer (the reflections are not an issue for the highly absorbing blackbody, but will be for other samples). The thermal-emission signal from the sample was collected by a parabolic mirror, sent into a moving-mirror Michelson interferometer, and then detected by a liquid-nitrogen-cooled mercury-cadmium-telluride (MCT) detector. The collecting parabolic mirror has a small numerical aperture (~0.05). In our setup, the beam path between the interferometer and detector includes several mirrors and apertures ["Optical components" in Fig. 1(a)].

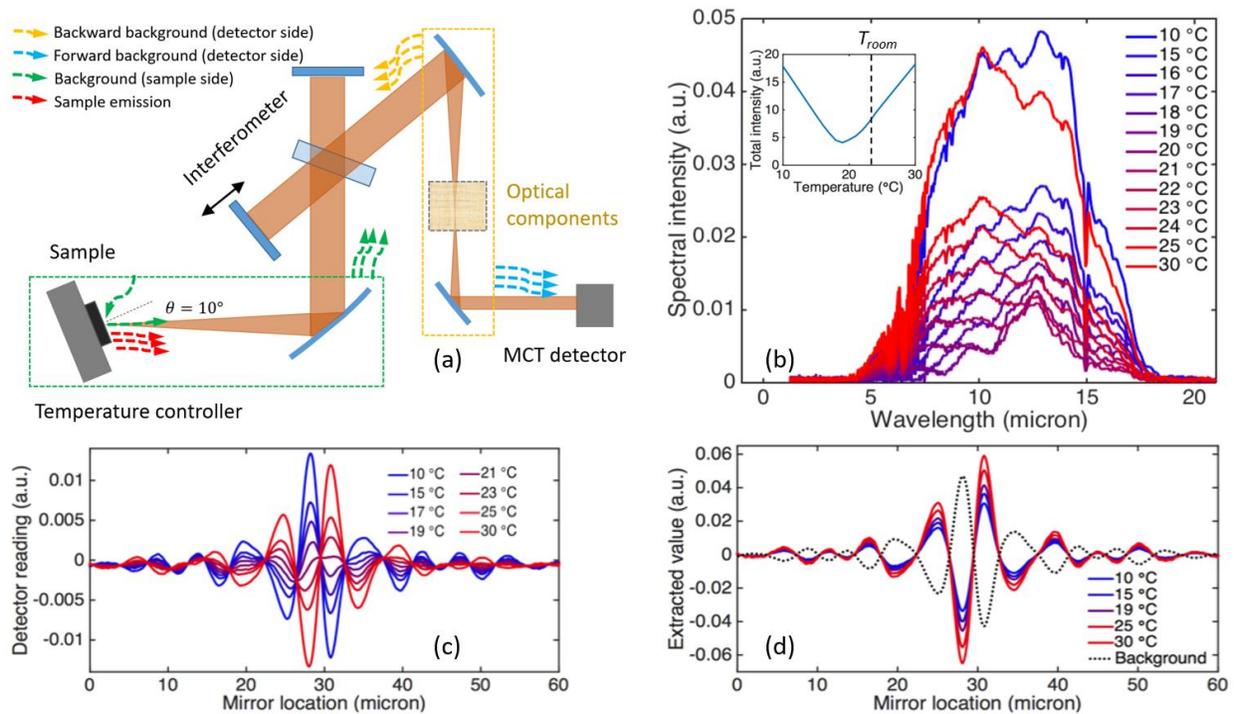

FIG. 1. (a) The optical path of our FTIR spectrometer. Light emitted from the sample (red arrows) is collected by a parabolic mirror, traverses the interferometer, passes through several optical components (mirrors and apertures), and is then sent to a liquid-nitrogen-cooled MCT detector. Most of the instrument components are at approximately room temperature, which results in several sources of background emission in both the forward (green and blue arrows) and backward (yellow arrows) directions. (b) Measured uncalibrated spectral intensity from the carbon nanotube (CNT) laboratory blackbody from 10 to 30 °C, which is minimized near 19 °C and increases when the sample is either heated or cooled. Inset: the integrated (but still uncalibrated) total spectral power over all wavelengths. The temperature of the room housing the experiment, $T_{room}$, is 23 °C. (c) Selected interferogram traces corresponding to the spectra in (b). The amplitude of the interferogram trace is minimized when sample temperature is near 19 °C, and there is a π phase shift between the traces when the sample temperature is above (red lines) and below (blue) 19 °C. (d) Extracted contributions from the sample (solid lines) and the combined background (black dotted).

In Fig. 1(b), we plot the measured emission spectra of the CNT blackbody from 10 to 30 °C. Note that the system was not purged, resulting in strong water and carbon-dioxide absorption lines for wavelengths outside of the 8 - 14 $\mu$m atmospheric transparency window. The measurement was performed with the temperature of the room housing the instrument ($T_{room}$) set to 23 °C. As expected, the measured signal increases when the sample is heated above $T_{room}$. However, when the sample is cooled substantially below $T_{room}$, the measured signal also increases, appearing to contradict physical intuition and Planck's law. In our setup, the transition between these two distinct behaviors occurs when the sample temperature is around 19 °C. To the best of our knowledge, this unexpected experimental observation has not been previously discussed in the photonics and thermal-emitter-engineering literature, though it has been reported in the remote-sensing community [26].

In Fig. 1(c), we plot several of the experimental interferogram traces that were Fourier-transformed to obtain the spectra shown in Fig. 1(b). Note that parts of the interferogram can be negative because our detector is AC-coupled. We observe that the interferogram traces for temperatures above and below 19 °C are roughly out of phase. Like the resulting spectrum, the magnitude of the interferogram is minimized around 19 °C.

**Interferogram analysis**

In most FTIR spectrometers, including our own, all components except for the sample and the detector are kept at approximately room temperature. Therefore, the non-negligible thermal emission from the various components of the instrument and the surrounding environment (together, the "background") must be accounted for during data analysis. In particular, there can be background emission from components located both before and after the interferometer. The emission from components located before the interferometer [*e.g.,* emission from the walls of the sample compartment that is reflected or scattered by the sample; green arrows in Fig. 1(a)] shares the same optical path as the sample emission: both are modulated via transmission through the interferometer and then received by detector. This part of the background adds to the total detected signal. The forward-emitting background [blue arrows Fig. 1(a)] from components located after the interferometer is transmitted directly to the detector without modulation, and thus does not contribute to the detected signal. Finally, the backward-emitting background from components located between the interferometer and the detector [yellow arrows, Fig. 1(a)] is modulated via reflection from the interferometer and then received by the detector. As explained in the following, this contribution is apparently subtracted from the total detected signal.

To understand the origin of this "negative" background signal, it is helpful to separate the experimental interferogram, $I(x, T)$, into a sum of the temperature-dependent emission from the sample, $I_S(x, T)$, and a temperature-independent background, $I_B(x)$:

$$I(x,T) = I_S(x,T) + I_B(x) \tag{1}$$

In Eq. 1, $x$ is the location of the moving mirror in the interferometer, $T$ is the sample temperature, and $I_B(x)$ represents the total background signal, which is due all of the contributions indicated by yellow, blue, and green arrows in Fig. 1(a). Note that $I_B$ can also become temperature dependent if the sample reflectivity is temperature dependent—via background emission reflected or scattered by the sample into the beam path—though this is not the case here. Using Eq. 1, these two components can be extracted from the experimental interferogram (Supplemental Material, Sect. 2), and are plotted in Fig. 1(d). In our experiment, the amplitude of the sample interferogram increases monotonically with increasing sample temperature, and the background interferogram is out of phase with the sample interferogram (note that we are referring to *interferogram phase*, rather than the phase of the electromagnetic waves). Due to this $\pi$ phase shift, the combined interferogram—and the resulting emission spectrum—is effectively representative of the difference between the two contributions, rather than the sum. When the sample temperature is high, the sample emission dominates, and the total interferogram is close to that of the sample alone. Conversely, background emission dominates when the sample temperature is below room temperature. The transition temperature at which the total interferogram is minimized is close to room temperature.

To confirm that the negative background indeed comes from background emission of room-temperature components, we performed thermal-emission measurements at several different temperatures of the room housing the experiment. Increasing the room temperature should result in an increase of the magnitude of the background emission, and thus the temperature at which the measured integrated spectral power is minimized. The experimental results agree with this prediction (Supplemental Material, Sect. 3).

**Analysis of the interferometric phase**

The $\pi$ phase shift of the background interferogram arises from the reflection of the backward-emitting background radiation by the interferometer, in contrast to the forward-emitting radiation from the sample and background, which is modulated by the interferometer via transmission. The differences in the phase shifts imparted by Michelson interferometers in reflection and transmission modes were discussed in the early 1990s [27], [28], and an explanation of this effect assuming a simple Fabry-Perot model for an FTIR beam splitter was given in ref. [28]. However, we are unaware of general discussions on this topic that do not assume a specific type of beam splitter. Here, we provide a simple, general explanation for the $\pi$ phase shift that is agnostic to the details of the interferometer except for the assumption that the beam splitter is not too lossy.

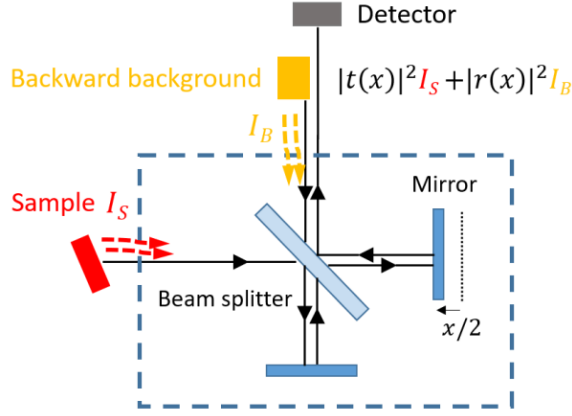

FIG. 2. Interaction of two thermal emitters ("Sample" and "Backward background") with an interferometer. The sample emission is modulated via transmission through the interferometer, whereas the backward background is modulated via reflection. The signal received by the detector is the sum of the two contributions.

The interaction of a Michelson interferometer with two emitters ("Sample" and "Backward background") is schematically shown in Fig. 2. Like in the experiment, the sample is rotated with a non-zero angle to the interferometer path so that multiple reflections need not to be considered. The propagation of light in the forward and backward directions through the interferometer can be defined by transmission and reflection coefficients. To simplify the analysis, we assume that the interferometer is lossless (*i.e.*, no absorption or scattering at the beam splitter and mirrors) and reciprocal (*i.e.*, the transmission is identical in both directions [29]). The reflection and transmission coefficients, $R(x)$ and $T(x)$, respectively, depend on the position of the moving mirror, $x$. The sample emission, $I_S$, is transmitted through the interferometer and received by the detector, and the backward-emitting background, $I_B$, is reflected by the interferometer towards the detector. The total signal received by the detector is therefore: $T(x)I_S + R(x)I_B$. For a lossless interferometer (*i.e.*, $T(x) + R(x) = 1$), the detected signal becomes:

$$I(x) = T(x) \cdot (I_S - I_B) + I_B. \tag{2}$$

Note that in Eq. 2, the second term, "$I_B$", is not seen by an AC-coupled detector because it is independent of $x$. Therefore, the detected signal becomes $T(x) \cdot (I_S - I_B)$. Thus, when the sample emission is comparable to the background emission ($I_S \approx I_B$), which can be the case for near-room-temperature samples, the total measured interferogram signal can be small [$I(x) \approx 0$]. Though our analysis assumes a lossless interferometer, Eq. 2 is also a good approximation for an interferometer with finite loss (*i.e.*, if the beam splitter is partially absorbing), and explains the counterintuitive temperature-dependent interferogram traces in Fig. 1(c, d).

**Impact of the "negative" background**

For an FTIR spectrometer, the measured Fourier-transformed emission signal $S_\alpha(\lambda, T)$ from a sample $\alpha$ is often expressed as [22]–[24]:

$$S_\alpha(\lambda, T) = \eta(\lambda)[\epsilon_\alpha(\lambda) I_{BB}(\lambda, T) + B_\alpha(\lambda)], \tag{3}$$

where $\eta(\lambda)$, $\epsilon_\alpha(\lambda)$, $I_{BB}(\lambda, T)$, and $B_\alpha(\lambda)$ represent the system response, sample emissivity, blackbody radiation from Planck's law, and the background emission, respectively. Note that the background can be sample-dependent; for example, there can be background thermal emission from the sample side reflected or scattered by the sample into the beam path. This analysis assumes a single combined background, which can arise from a combination of sources before and after the interferometer; knowledge of the precise breakdown of each contribution may not be necessary. From Eq. 3, the emissivity of sample $\alpha$ can be obtained using a reference sample $\beta$ with known emissivity $\epsilon_\beta(\lambda)$ via:

$$\epsilon_\alpha(\lambda) = \epsilon_\beta(\lambda) \frac{S_\alpha(\lambda, T_1) - S_\alpha(\lambda, T_2)}{S_\beta(\lambda, T_1) - S_\beta(\lambda, T_2)}. \tag{4}$$

Note that Eq. 4 assumes the sample emissivity does not change significantly with temperature, which is the case for most thermal emitters. To obtain the emissivity using Eq. 4, emission measurements at two different temperatures, $T_1$ and $T_2$, can be used.

To illustrate the influence of this "negative" background on FTIR-based thermal-emission measurements, we measured thermal emission from a polished fused-silica wafer and a laboratory blackbody using the same setup, varying the temperature from 5 to 35 °C in steps of 1 °C (Supplemental Material, Sect. 4). No polarizer was used in this measurement. We selected $T_1$ and $T_2$ with a small interval of $T_2 - T_1 = 5$ °C, and then extracted the emissivity of silica using Eq. 4.

In Fig. 3(a), we plotted the experimental emissivity of the fused-silica wafer from 7 to 16 $\mu$m, extracted using Eq. (4) and data from different five-degree temperature ranges, within an overall range from 10 to 25 °C. For reference, we calculated the emissivity using Kirchhoff's law (green curve), using the optical properties of fused silica from variable-angle spectroscopic ellipsometry (VASE) and Fresnel coefficients (Supplemental Material, Sect. 5). The emissivity value extracted using emission data from $T_1 = 10$ °C and $T_2 = 15$ °C (blue curve) has a very large error near 9 and 13 $\mu$m. The value extracted using the emission data from $T_1 = 15$ °C and $T_2 = 20$ °C (red curve) has large errors at almost every wavelength, because this range includes the temperature at which the detected FTIR signal is minimized, $T_{transition} \sim 18$ °C. As shown in Fig. 1(b), when the sample temperature crosses $T_{transition}$, the system response function $\eta(\lambda)$ changes its sign: as the sample temperature is increased, the measured signal increases above $T_{transition}$ but decreases below $T_{transition}$. Clearly, Eq. 3 is no longer valid near $T_{transition}$. Therefore, one should always make sure $T_1$ and $T_2$ are both on the same side of $T_{transition}$ when applying Eq. 4. As the data collection temperature range is increased above 20 °C (yellow curve), the measured emissivity becomes close to the calculated value (green curve).

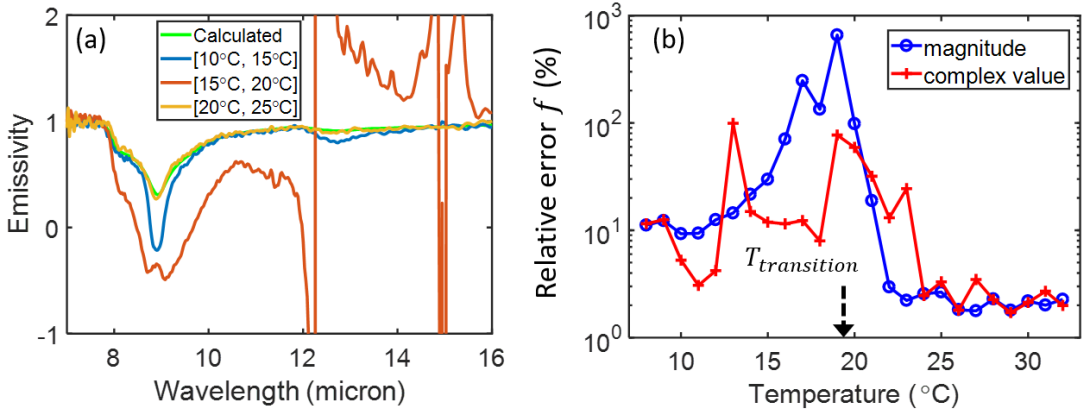

FIG. 3. (a) Measured polarization-averaged emissivity of a thick fused-silica sample at an angle of 10°, extracted from the magnitude of the measured spectrum (*i.e.*, the magnitude of the Fourier transform of the interferogram), from different temperature ranges. The corresponding calculation based on optical constants from spectroscopic ellipsometry is shown in green. (b) Averaged relative error (Eq. 5) in the extracted emissivity as a function of the central temperature at which the emission data was taken, using only the magnitude of the spectrum (blue) or the full complex value (red).

To quantify this effect, we calculated the emissivity $\epsilon_{exp}(\lambda, T)$ by gradually changing the central temperature $T$ ($T_1 = T - 2$ and $T_2 = T + 2$), and obtained the relative error using

$$f(T) = \frac{1}{\lambda_2 - \lambda_1} \int_{\lambda_1}^{\lambda_2} \frac{\epsilon_{exp}(\lambda, T) - \epsilon_c(\lambda)}{\epsilon_c(\lambda)} d\lambda, \tag{5}$$

where $\epsilon_c(\lambda)$ is the calculated emissivity [green curve, Fig. 3(a)], which we assume to be exact. The relative error $f$ is plotted in Fig. 3(b) for different central $T$, with $\lambda_1$ and $\lambda_2$ chosen to be 7 and 16 $\mu$m, respectively. In the same figure, $T_{transition}$, the temperature at which the measured emission signal is minimized, is identified. The large error near $T_{transition}$ is the result of the change of the system response of the FTIR spectrometer; $f$ decreases at both lower and higher temperatures. An asymmetry of $f$ respect to $T_{transition}$ is observed, which is related to the fact that the negative background does not fully cancel out the sample emission at $T_{transition}$, and the measured signal is not minimized at every wavelength at $T_{transition}$.

An alternative calibration method that was proposed in ref. [23] uses the full complex measured spectrum (*i.e.*, the Fourier transform of the interferogram) instead of using only the magnitude, as we did in Eqs. (3, 4). This complex spectrum can be expressed as:

$$\widetilde{S_\alpha}(\lambda, T) = \eta(\lambda)\big[\epsilon_\alpha(\lambda) I_{BB}(\lambda, T) + B_\alpha(\lambda) e^{i\phi_{0,\alpha}(\lambda)}\big] e^{i\phi_\alpha(\lambda)}, \tag{6}$$

where $\phi_{0,\alpha}$ is the background phase and $\phi_\alpha$ is the instrument-response phase, both for sample α. Then, the emissivity can be extracted using:

$$\epsilon_\alpha(\lambda) = \epsilon_\beta(\lambda) \left| \frac{\widetilde{S_\alpha}(\lambda, T_1) - \widetilde{S_\alpha}(\lambda, T_2)}{\widetilde{S_\beta}(\lambda, T_1) - \widetilde{S_\beta}(\lambda, T_2)} \right|. \qquad (7)$$

The calibration error using Eq. (7) is also plotted in Fig. 3(b). Compared to the use of Eq. (4), the error is reduced close to $T_{transition}$, where it is nevertheless very large using either method. For sample temperatures significantly above $T_{transition}$, Eq. (4) seems to be more robust.

Since the thermally radiated power from an emitter depends on its emissivity, the value of $T_{transition}$ depends on the emissivity of both the sample and the background. More specifically, $T_{transition}$ is expected to increase for an emitter with a lower emissivity. Therefore, such a negative background could have an impact on even relatively high-temperature thermal-emission measurement when the sample has a low emissivity.

**Further discussion and conclusion**

In principle, if one were to place the detector immediately after the interferometer, any backward-propagating background emission should be negligible. In practice, FTIR systems are often multifunctional, *e.g.*, set up for some combination of transmission, reflection, emission, and other measurements, possibly with removable or fixed accessories such as microscopes and custom stages. For example, our system has a number of emitting components that contribute to the background, including several aluminum mirrors, a prism for coupling of a laser pump, and apertures that restrict the area of detection on the sample. We believe that many FTIR-based systems have emitting components that may be impractical to remove, and must be accounted for, following the approach outlined above. Alternatively, a room-temperature detector can be utilized, which would then be in equilibrium with all of the background sources. This approach, however, can result in low signal-to-noise ratios [30], and requires that the detector temperature is identical to ~~that of~~ that of the background sources.

If the emissivity of the sample does not change with temperature, one can avoid the impact of the negative background by measuring its thermal emission at a higher temperature and obtain accurate results using Eqs. (3-4). However, there are some cases that one must perform low-temperature measurements (*e.g.*, the experiments in refs. [6], [20], [21]). If the target thermal-emission temperature is close to $T_{transition}$, one solution is to decrease the ambient temperature to lower $T_{transition}$.

Further complications occur for thermal emitters near room temperature whose emissivity is temperature-dependent (*e.g.*, those in refs. [31], [32], [33]). In this case, one need not only to decrease the ambient temperature, but also to choose $T_1$ and $T_2$ that are close enough to each other such that the emissivity is not very different at the two temperatures; this approach can result in low signal-to-noise ratios. Finally, for non-equilibrium or non-reciprocal systems where Kirchhoff's law does not apply, the background must be fully calibrated to obtain the actual thermal-emission signal.

We note that the peculiarities described in this manuscript can be avoided by using dispersive spectrometers. However, the "multiplex advantage" and other features of FTIRs make them otherwise ideal for various mid-infrared measurements, especially for situations with weak signal (*e.g.*, low-temperature thermal emission) or narrowband features (see the discussion on narrowband emitters in Ref. [5]).

To summarize, in this work we observed counterintuitive phenomena when performing near-room-temperature thermal-emission measurements using an FTIR spectrometer, *i.e.*, an apparent "negative" background signal and a sign inversion of the system response function. We found that these features arise from the background emission of optical components positioned between the interferometer and the detector. Similar behavior has been previously discussed in remote-sensing literature, but not in the nanophotonics and thermal-emission-engineering communities. We expect these and related issues to be significant in different kinds of FTIR systems, and provided suggestions and guidance on low-temperature thermal-emission measurements.

The authors acknowledge discussions with Romain Blanchard from Pendar Technologies and Daniel Wasserman from University of Texas - Austin. This work is supported by the Office of Naval Research (N00014-16-1-2556) and the Department of Energy (DE-NE0008680).

Supplemental Material:

# Thermal-emission measurements near room temperature using Fourier-transform infrared spectroscopy

Yuzhe Xiao[1], Alireza Shahsafi[1], Chenghao Wan[1,2], Patrick J. Roney[1], Graham Joe[1], Zhaoning Yu[1,3], Jad Salman[1], and Mikhail A. Kats[1,2,3*]

[1]*Department of Electrical and Computer Engineering, University of Wisconsin-Madison, Madison, Wisconsin, USA 53706*
[2]*Materials Science and Engineering, University of Wisconsin-Madison, Madison, Wisconsin, USA 53706*
[3]*Department of Physics, University of Wisconsin - Madison, Madison, Wisconsin 53706, USA*
*\*Email address: mkats@wisc.edu*

## 1. Response function of our FTIR spectrometer

To separate the background and sample interferogram from the total interferogram (Sec. 2), we must measure the spectral response function of the FTIR spectrometer, $\eta(\lambda)$, which includes the detector responsivity and the transmission efficiency of the optical system. We obtain this response function by measuring thermal emission from a laboratory blackbody at different temperatures well above room temperature (50 and 100 °C, Fig. S1(a)), and using Eq. (3) in the main text:

$$\eta(\lambda) = \frac{S_{BB}(\lambda,T_1) - S_{BB}(\lambda,T_2)}{\epsilon_{BB}[I_{BB}(\lambda,T_1) - I_{BB}(\lambda,T_2)]} \tag{S1}$$

where $S_{BB}(\lambda, T)$ is the measured Fourier-transformed signal from the laboratory blackbody of emissivity $\epsilon_{BB} \cong 0.98$. The resulting normalized system response function is plotted in Fig. S1(b). Note that the system is not purged, resulting in strong water and carbon-dioxide absorption lines for wavelengths outside of the 8 - 14 $\mu$m atmospheric transparency window.

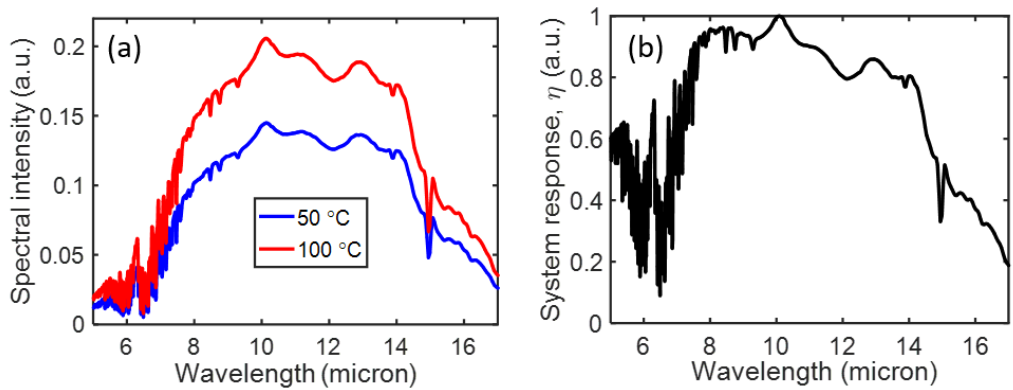

FIG. S1. (a) Measured thermal-emission spectra at 50 and 100 °C for a laboratory blackbody. (b) Normalized extracted system response $\eta(\lambda)$.

## 2. Extraction of signal and background interferogram

In this section, we show how we extract the contributions of the sample emission and the background from the total interferogram. The measured interferogram traces for a laboratory blackbody with temperature varying from 10 to 30 °C are plotted in Fig. S2(a). As discussed in the main text, the total interferogram consists of two contributions:

$$I(x,T) = I_S(x,T) + I_B(x) \tag{S2}$$

Since the background contribution $I_B(x)$ does not change with sample temperature $T$, we can eliminate the background component by subtracting two curves with different $T$. In Fig. S2(b), we plot the differences of pairs of interferogram traces with temperatures differing by of 1 °C: $I(x, T+1) - I(x, T)$, for $T$ from 15 to 24 °C.

As shown here, the difference seems to be a constant that does not change with respect to $T$. This difference comes from the increased thermal emission signal from the sample due to the increased $T$, which can be modeled theoretically. As shown below, the difference is expected to be approximately independent of $T$ because the small window for $T$ (15 to 24 °C).

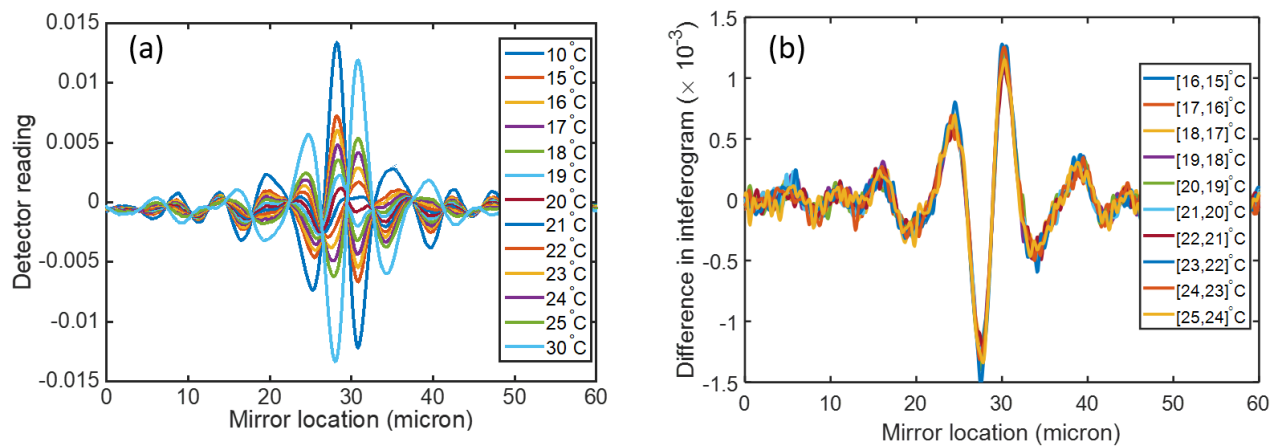

FIG. S2. (a) Experimental interferogram traces for a laboratory blackbody from 10 to 30 °C. (b) Difference in pairs of experimental interferogram traces with a temperature difference of 1 °C.

Note that the sample interferogram is obtained by taking the interference between sample thermal emission from two paths (along the two arms of the interferometer) [S1]:

$$I_S(x,T) \sim \int_0^\infty I_{BB}(T,\lambda)\,\eta(\lambda)\epsilon_s(\lambda)\left|1+e^{-i4\pi x/\lambda}\right|^2 d\lambda, \tag{S3}$$

where $\eta(\lambda)$ is the system response, $\epsilon_s(\lambda)$ is the sample emissivity, $I_{BB}(T,\lambda)$ is the blackbody radiation from Planck's law [S2], and the integration is performed over all wavelength. Note that ideally, the interferogram is a symmetric function of $x$, as can be seen from Eq. S3. In practice, interferometers can have path differences that change with frequency, leading to an asymmetric interferogram [S3].

A similar equation can be written for the contribution of the background. The maximum value of the sample interferogram (zero delay between two branches of the interferometer, $x=0$) is:

$$I_S(0,T) \sim \int_0^\infty I_{BB}(T,\lambda)\,\eta(\lambda)\epsilon_s(\lambda)d\lambda. \tag{S4}$$

Therefore, the ratio for the maximum signal at two different temperatures $T_1$ and $T_2$ is:

$$\rho_{T_1/T_2} = \frac{\int_0^\infty I_{BB}(T_1,\lambda)\eta(\lambda)\epsilon_s(\lambda)d\lambda}{\int_0^\infty I_{BB}(T_2,\lambda)\eta(\lambda)\epsilon_s(\lambda)d\lambda}. \tag{S5}$$

For a sample with wavelength-independent emissivity $\epsilon_s(\lambda) = \epsilon_s$ and a system with wavelength-independent response $\eta(\lambda) = \eta_0$, $\rho_{T_1/T_2} = T_1^4/T_2^4$ according to the Stefan-Boltzmann law. In our case, the sample (a laboratory blackbody) has very uniform emissivity across all wavelengths, but the FTIR system response has a finite bandwidth. Using the system response function $\eta(\lambda)$ calibrated in Fig. S1(b), we calculate the expected ratio of the interferogram magnitude from the sample at different temperatures and summarize them in Table S1.

Table S1. Calculated ratio of the sample interferogram peak values at different temperatures. Also shown are the ratios predicted by Stefan-Boltzmann's law. The temperature value $\gamma$ is chosen from 15 to 24 °C.

| $\rho_{T_1/T_2}$ | [15, 10] °C | [$\gamma$+1, $\gamma$] °C | [30, 25] °C |
|---|---|---|---|
| Eq. S4 | 1.187 | ~1.033 | 1.174 |
| Stefan-Boltzmann law | 1.073 | ~1.014 | 1.069 |

Note that the ratio from different temperatures $\gamma$ is approximately the same because of the small temperature window (15 to 24 °C). Also note that the ratios from the Stefan-Boltzmann law are slightly smaller due to the finite bandwidth of our FTIR [Fig. S1(b)], *i.e.*, the span of integral is finite. Knowing the ratio of the sample interferograms at different temperatures, we can extract the individual contributions from the sample and background as follows. We assume the sample interferogram at 15 °C is $f(x)$:

$$I_S(x, 15 \,°C) = f(x). \tag{S6}$$

The sample interferogram at 16 °C can be obtained using $f(x)$ and the ratio calculated in Table S1:

$$I_S(x, 16 \,°C) = 1.033 \, f(x) \tag{S7}$$

Therefore, difference of the total interferogram at 15 and 16 °C is

$$I(x, 16 \,°C) - I(x, 15 \,°C) = I_S(x, 16 \,°C) - I_S(x, 15 \,°C) = 0.033 \, f(x). \tag{S8}$$

Note that the background contribution $I_B(x)$ is cancelled out in Eq. S8. With the experimental data shown in Fig. S2 (b), $f(x)$ (the sample interferogram at 15 °C) can be determined. Once $I_S$ at one temperature is known, its value at all of the other temperatures can be deduced through the relation in Table S1. Thus $I_B(x)$ can be obtained through

$$I_B(x) = I(x, T) - I_S(x, T). \tag{S9}$$

The extracted contributions from the sample $I_S(x, T)$ and the background $I_B(x)$ are plotted in Fig. S3(a) and (b), respectively. As shown in Fig. S3(b), the values of $I_B(x)$ extracted using Eq. S9 from different $T$ are the same, as expected. The magnitude of $I_S(x, T)$ increases with $T$.

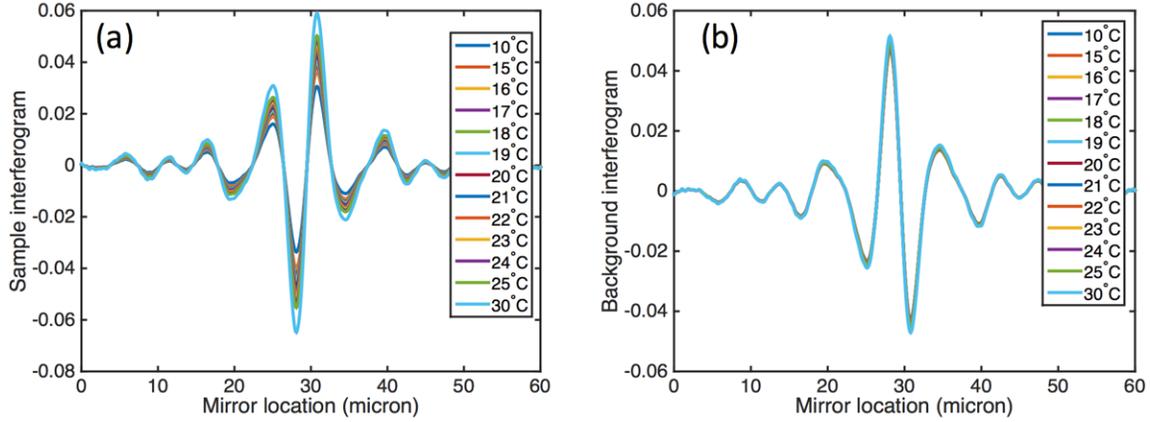

FIG. S3. Extracted (a) sample and (b) background interferogram traces from the total interferograms in Fig. S2.

## 3. Emission measurements with different ambient temperatures

The magnitude of background emission increases as the ambient temperature is increased. Therefore, the corresponding temperature at which the background neutralizes the sample signal in our experiment should increase as well, since our background emission is dominated by the backward-emitting signal from the optical components of our FTIR. To confirm this, we performed near-room-temperature measurements of the fused-silica wafer used in the main text, but changed the temperature of the room housing the experiment. The measured spectra of the fused-silica wafer from 10 to 30 °C with the ambient temperature set to 22, 24, and 25 °C are plotted in Fig. S4(a-c), respectively.

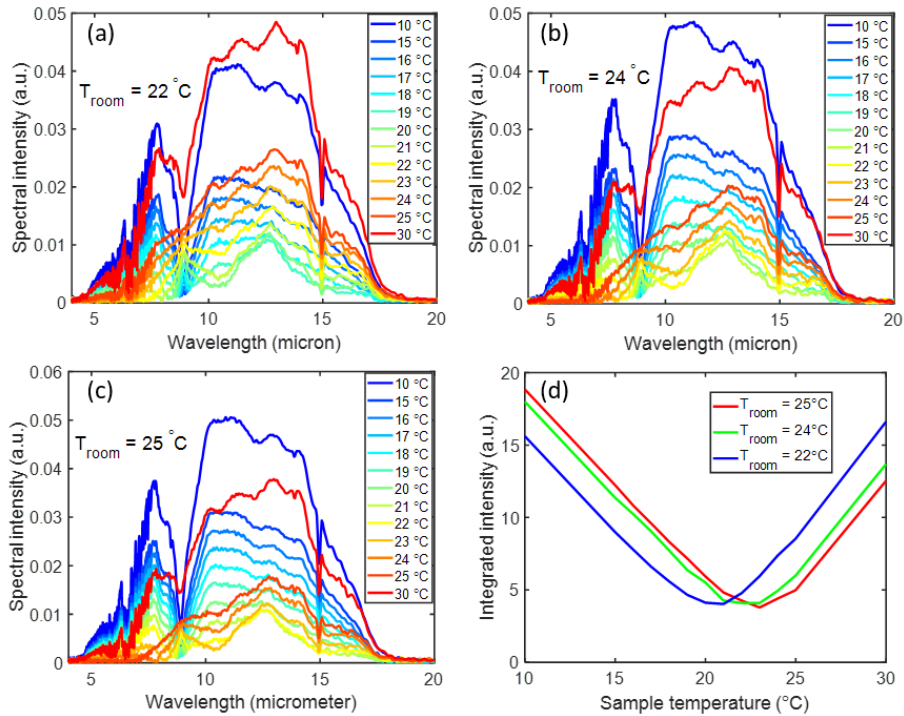

FIG. S4. Measured emission spectra from a fused-silica wafer from 10 to 30 °C with the temperature of the room housing the experiment set to (a) 22, (b) 24, and (c) 25 °C. (d) The total integrated power as a function of the sample temperature at different ambient temperatures.

As shown in Fig. S4(a), the measured spectrum from 10 to 15 $\mu$m has its minimum value at a sample temperature of 20 °C, when the ambient temperature was set to 22 °C. When the ambient temperature is increased up to 24 °C (Fig. S4-b) and 25 °C (Fig. S4-c), the corresponding minimized signal is at 22 and 23 °C, respectively. To better observe this phenomenon, we integrated the spectra across the wavelengths between 5 $\mu$m to 20 $\mu$m at the three ambient temperatures (Fig. S4-d). We see a clear increase of the transition temperature as the lab-room temperature is increased.

Note that the dip in the measured signal for fused silica near 9 µm is due to the corresponding dip in its emissivity close to its phonon resonance. The total measured emission contains both the sample emission and the background. If the temperature difference between the sample and the background is large enough (i.e., the sample emission magnitude is much larger than that of the background), the measured data is more like the emission from the sample, which has the corresponding emissivity feature. In the case of fused silica, this feature is a dip near 9 µm. When the difference between the temperatures of the sample and the background is small enough, the data becomes mixed. As shown in Fig. S4, the dip is obvious only for temperature not too close to the transition temperature. This is also seen from Fig. S4, where the 30 °C dip is largest in (a) and smallest in (c).

If the background emission were to completely cancel out the sample emission at all wavelength at the transition temperature, one would expect the same spectrum of the sample at two different temperatures that are roughly equal "distance" from the transition temperature (for small-enough distances). In our case, we do not have total cancellation. Therefore, there is an asymmetry in the measured spectrum vs. temperature on two sides of the transition temperature.

4. **Thermal emission data for calibrating the emissivity of our fused-silica wafer**

The thermal emission from a fused-silica wafer and a laboratory blackbody were measured using our FTIR with temperature increasing from 5 to 35 °C. Both samples were rotated by 10° with respect to the beam path. No polarizer was used, so the measured signal was the averaged value for p and s polarization. The emission data shown in Fig. S5 are used to obtain the emissivity of the fused-silica wafer.

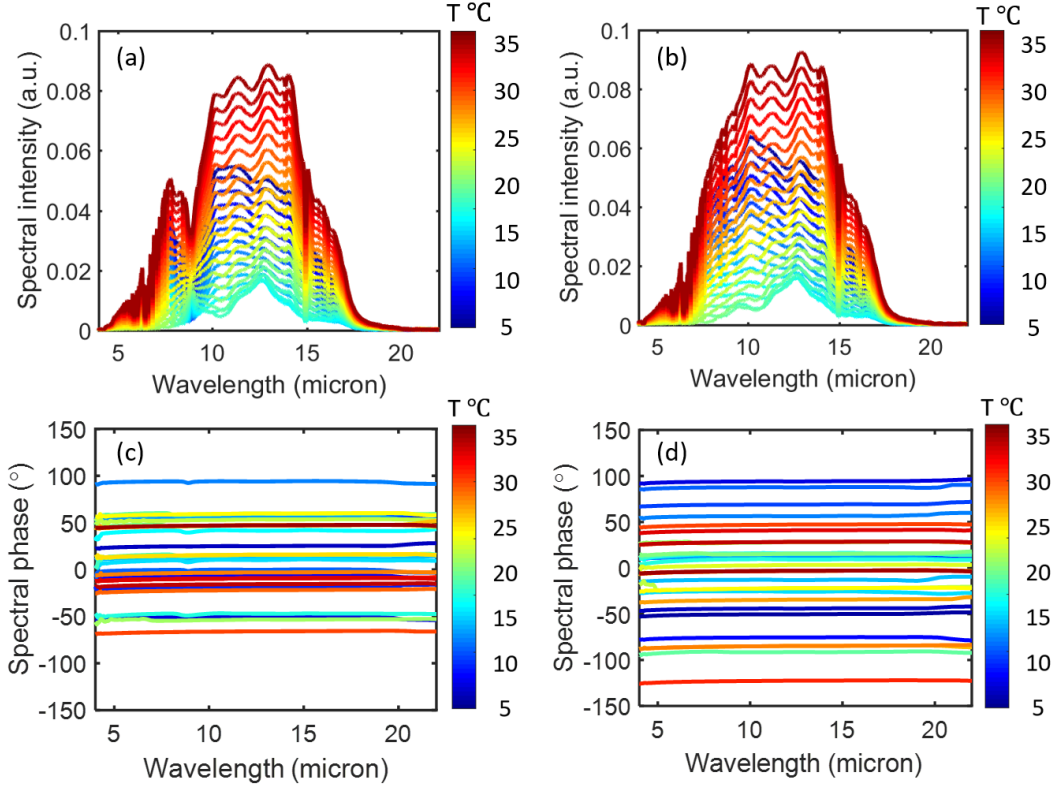

FIG. S5. Measured the magnitude of the thermal-emission spectrum from (a) a fused-silica wafer and (b) a laboratory blackbody ($\epsilon \approx 0.98$) from 5 to 35 °C in step of 1 °C. (c) and (d) shows the corresponding phase of the spectrum for fused-silica and laboratory blackbody.

## 5. Emissivity of our fused-silica wafer from ellipsometry

Ellipsometry measurements were performed on our polished fused-silica wafer with incident angles of 35, 45 and 55°, for free-space wavelengths from 4 to 20 $\mu$m. The complex refractive indices were then extracted by fitting the raw data ($\Psi$ and $\Delta$), with the results shown in Fig. S6(a). Figure S6(b) shows the calculated polarization-averaged emissivity at the incident angle of 10°, obtained by Kirchhoff's law for a flat opaque emitter: $\epsilon(\lambda) = \alpha(\lambda) = 1 - R(\lambda)$. The reflectivity, $R(\lambda)$, is the averaged value for s- and p-polarization at 10°, obtained by Fresnel equation using the material data shown in (a).

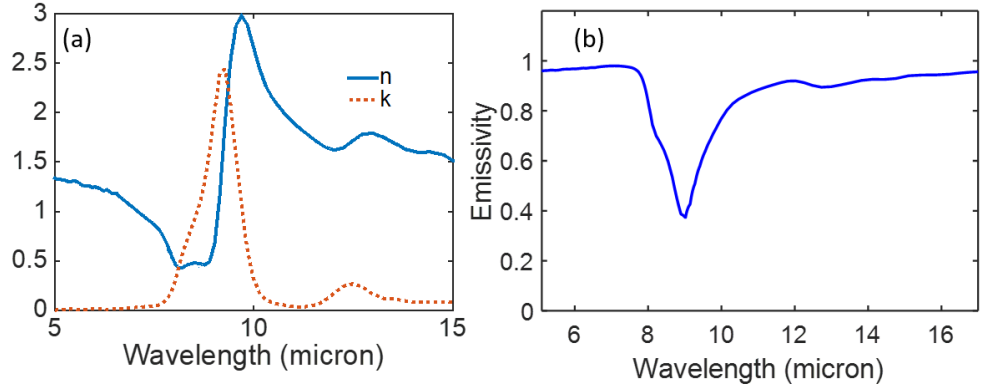

FIG. S6. (a): Real (solid) and imaginary (dotted) part of the refractive index of our fused-silica wafer, extracted using spectroscopic ellipsometry. (b): Calculated polarization-averaged emissivity at an incidence angle of 10°, using the material properties in (a).

## Supplementary references